# The Moon as a Recorder of Nearby Supernovae

**Ian A. Crawford**

Department of Earth and Planetary Sciences, Birkbeck College, University of London

i.crawford@bbk.ac.uk

## Abstract

The lunar geological record is expected to contain a rich record of the Solar System's galactic environment, including records of nearby (i.e. ≤ a few tens of parsecs) supernova explosions. This record will be composed of two principal components: (i) cosmogenic nuclei produced within, as well as radiation damage to, surface materials caused by increases in the galactic cosmic ray flux resulting from nearby supernovae; and (ii) the direct collection of supernova ejecta, likely enriched in a range of unusual and diagnostic isotopes, on the lunar surface. Both aspects of this potentially very valuable astrophysical archive will be best preserved in currently buried, but nevertheless near-surface, layers that were directly exposed to the space environment at known times in the past and for known durations. Suitable geological formations certainly exist on the Moon, but accessing them will require a greatly expanded programme of lunar exploration.

## 1. Introduction

In contrast to the geologically active surface of the Earth, which has been protected by a dense atmosphere and strong magnetic field throughout Solar System history, the surface of the Moon has been exposed directly to the space environment since the Earth-Moon system formed approximately 4.5 billion years ago. During this time the Solar System has orbited the Galaxy approximately twenty times and will have experienced a wide range of galactic environments, including passages through spiral arms and star-forming regions where the rate of supernova (SN) explosions will have been enhanced. This raises the possibility that such events have been recorded in the lunar geological record, and especially in the surface regolith which is, or has been, exposed directly to space. Such a record could take several forms, including radiation damage to near surface materials exposed to an enhanced flux of relativistic subatomic particles (i.e. cosmic rays), the production of cosmogenic nuclei when such particles interact with target nuclei in near-surface materials, and, for nearby events, the possibility of direct collection of SN ejecta, likely bearing unique isotopic signatures, on the lunar surface. As argued below, the geological evolution of the Moon, having been relatively inactive for billions of years, but just active enough to occasionally cover, and thereby protect, surficial deposits bearing evidence of the outer space environment, makes the lunar geological record an ideal recorder of such events throughout Solar System history.

## 2. The galactic environment of the Solar System

The Solar System has orbited the centre of the Milky Way Galaxy approximately twenty times since the Sun formed 4.6 billion years ago (e.g. Gies and Helsel, 2005; Overholt et al., 2009). During this time the Solar System will therefore have been exposed to a wide range of galactic environments, including passing through spiral arms where an enhanced probability of passing close to SN explosions, and associated supernova remnants, would be expected. In searching for a record of such events in the Solar System it is actually the Sun's motion with respect to the Galaxy's spiral arm structure that is important. There is considerable uncertainty regarding the angular velocity of the Sun with respect to the more slowly rotating pattern of spiral arms, with values between 0 and 13.5 km s$^{-1}$ kpc$^{-1}$ appearing in the literature; see Shaviv (2003) for a summary of the earlier literature, and Overholt et al. (2009) for a more recent discussion. The current consensus appears to be that the value lies probably somewhere between 6 and 13.5 km s$^{-1}$ kpc$^{-1}$ (Overholt et al., 2009), which correspond to orbital periods relative to the spiral pattern of approximately 1 Gyr and 450 Myr, respectively. These values imply that the Solar System will have traversed the entire spiral pattern of the Galaxy, which appears to consist of four main spiral arms and a number of inter-arm 'spurs' (Fig. 1; Churchwell et al., 2009), between 5 and ten times in the course of its history.

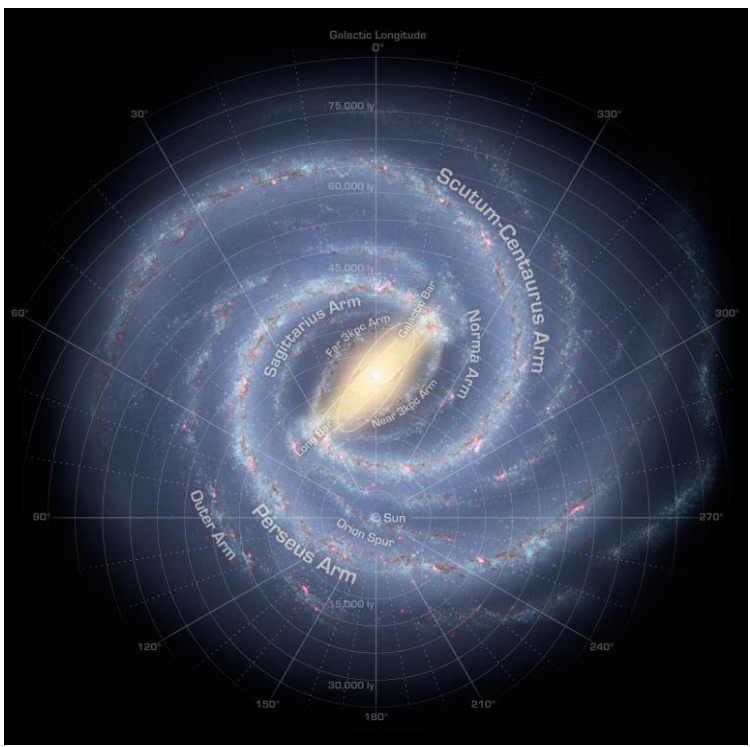

**Figure 1.** A recent reconstruction of the structure of the Milky Way Galaxy (Churchwell et al., 2009). Galactic longitudes, major spiral features, and the present location of the Sun are indicated. Since the Solar System formed it will have passed through the entire spiral pattern between five and ten times (depending on the assumed angular velocities of the Sun and the Galaxy's spiral pattern; see text), and will therefore have experienced a wide range of different galactic environments. Evidence for this cosmic odyssey may be preserved in the lunar geological record. (Image credit: NASA/JPL-Caltech/R. Hurt/Wikipedia Commons).

Whenever the Sun passes through a spiral arm several observable consequences may be expected. Of most relevance to this chapter is an enhanced flux of galactic cosmic rays (GCRs) resulting from the fact that massive stars, which comprise the progenitors of Type II supernovae, are largely confined to the spiral arms, and that supernova remnants are major sources of cosmic rays (e.g. Ackermann et al., 2013). Shaviv (2003) has attempted to model this effect and finds that the GCR flux during spiral arm passages may be enhanced by factors of between two and five compared to inter-arm regions. In addition to records of nearby supernovae, spiral arm passages may also be expected to result in compression of the heliosphere owing to a denser interstellar medium (which would also result in an increased GCR flux), increased accretion of interstellar dust particles onto planetary surfaces, and perhaps an enhanced cometary impact rate on the terrestrial planets (e.g. Shaviv, 2006; see also discussion by Crawford et al., 2010, and references cited therein).

There have been multiple attempts (e.g. Shaviv 2003; Filipovic et al., 2013; see also the summary by Overholt et al., 2009) to use the Earth's geological record, and especially the alleged periodicity of mass extinction events and/or ice ages, to constrain models of galactic structure. Overholt et al. (2009) have shown that to-date no such correlations with galactic structure have been convincingly demonstrated and, given all the uncertainties in interpreting the Earth's complex geological and biological history, this is hardly surprising. However it is important to realise that much better records of spiral arm passages probably exist elsewhere in the Solar System. Indeed, all the effects of spiral arm passages noted above have the potential to leave records in the in the near surface environments of geologically inactive and airless bodies such as asteroids and the Moon. Uncovering them is a major potential *astrophysical* benefit of continued Solar System exploration (e.g. Spudis, 1996; Crawford and Joy, 2014).

Although using Solar System records to infer how the Sun's galactic environment has changed over one or more revolutions of the Galaxy will be a challenging project for future research, we do at least have reliable information on the present galactic environment of the Sun. It has been known for several decades that the Sun is currently passing through a hot ($T$~$10^6$ K), mostly ionised, low density ($n_H$ ~ 0.005 cm$^{-3}$) region of the interstellar medium of the order of 150 pc in size that has become known as the Local Bubble (e.g. Frisch et al., 2011; Galeazzi et al., 2014). The origin of the Local Bubble has been a subject of debate ever since its discovery, but there is now a consensus that multiple (>10) supernova events arising within the nearby Sco-Cen OB Association during the last 10-15 Myr were largely responsible (Maíz-Apellániz, 2001; Breitschwerdt et al., 2009; 2016). The Sun's velocity of ~20 kms$^{-1}$ (~20 pc Myr$^{-1}$) relative to the local standard of rest implies that the Solar System will have resided within the Local Bubble for several (probably ≥ 3) million years and may therefore have experienced an enhanced GCR flux during this time.

## 3. Possible supernova records on the Moon

Studies of Apollo samples have revealed that the lunar regolith (Fig. 2) is efficient at collecting and retaining materials that impinge upon it from space. Specifically, in addition to meteoritic debris, which is primarily of interest to planetary scientists studying the evolution of the Solar System, the regolith also contains a physical record of a range of important astrophysical processes. These include the flux and composition of the solar wind, as well as cosmogenic nuclides and tracks of radiation damage produced by high energy GCRs (e.g. Crozaz et al., 1977; McKay et al., 1991; Lucey et al., 2006). The GCR records are of particular relevance for SN studies because, as discussed above, nearby supernovae will result in a temporal enhancement in the cosmic ray flux which may therefore be recorded in the lunar geological record. In addition, if a SN explosion were to occur within a few tens of parsecs of the Solar System (e.g. those thought to be responsible for creating the Local Bubble) there is the additional possibility that supernova ejecta might be directly implanted onto the lunar surface (e.g. Benítez et al., 2002; Fry et al., 2015). These two broad classes of possible SN records on the Moon are discussed below.

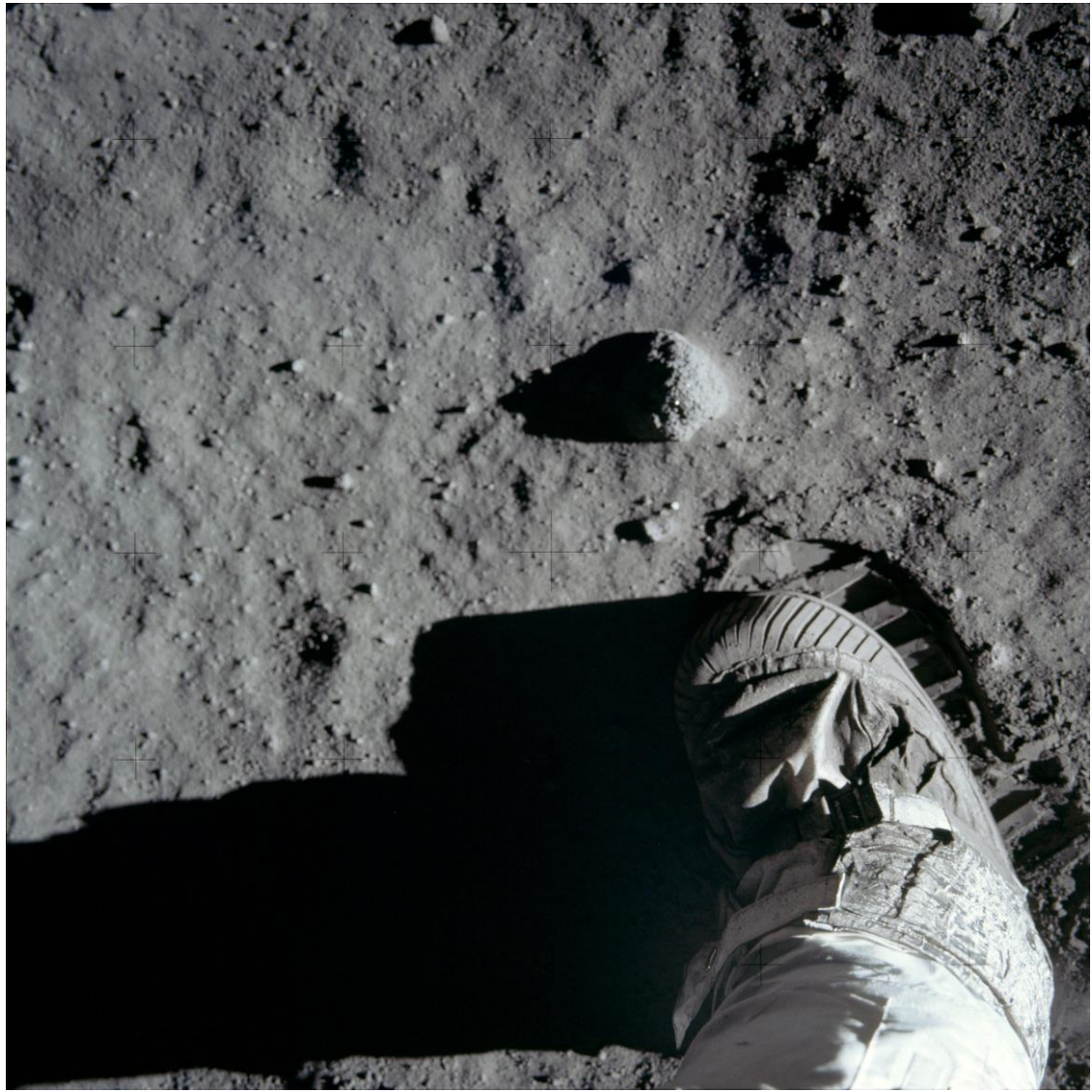

**Figure 2.** Close-up of the lunar regolith with astronaut's boot for scale. The uppermost metre or so of the regolith is an efficient recorder of materials and influences impinging on it from space, including meteoritic debris, solar wind particles, cosmogenic nuclides produced by GCRs, and possibly ejecta from nearby supernova explosions. See text for discussion. (NASA image AS11-40-5880).

3.1 Detecting and enhanced GCR flux from nearby supernovae

When GCRs interact with atoms in geological materials a variety of cosmogenic nuclei are produced as a result of spallation and neutron capture reactions (e.g. Eugster et al., 2006; Wieler et al., 2013). Typical cosmogenic nuclei include the stable isotopes $^{3}$He, $^{21}$Ne, $^{38}$Ar, $^{83}$Kr and $^{126}$Xe, and the unstable isotopes $^{10}$Be, $^{36}$Cl, and $^{39}$Ar, and it is by measuring the concentrations of these and other cosmogenic isotopes, under the assumption of a constant (or at least known) background GCR flux, that cosmic ray exposure (CRE) ages are commonly derived for lunar and planetary materials (Eugster, 2003; Wieler et al., 2013).

In reality, however, for the reasons discussed in Section 2, the GCR flux must vary with the changing galactic environment of the Solar System. In this context, it is interesting to note that studies of iron meteorites (Lavielle et al., 1999; Marti et al., 2004) have indicated possible variations in the GCR flux based on discrepancies between CRE ages obtained from the concentrations of stable cosmogenic nuclei and radioactive nuclides having different half-lives (e.g. $^{36}$Cl/$^{36}$Ar, $^{39}$Ar/$^{38}$Ar, $^{10}$Be/$^{21}$Ne). For example, Marti et al. (2004) inferred that the primary GCR flux over the last 10 Myr may have been almost 40% higher than the average of the period 150-700 Myr ago. The GCR flux over the latter period will have been averaged over a significant fraction of the galactic disc, and an enhancement within the last 10 Myr could be consistent with the Sun's encounter with the local Orion Spur (Fig. 1) and, more particularly, the Local Bubble. More recently, however, Wieler et al. (2013) have reanalysed the meteoritic data and have not been able to confirm the evidence for an enhanced GCR flux within this time period.

As stressed by Wieler et al. (2013), the usefulness of meteorites in attempting to identify variations in the flux of GCRs is constrained, as it is for all extraterrestrial samples studied to-date, by the fact that they can only record an integrated GCR flux since they became exposed to the space environment. For this reason, Wieler et al. combined their meteoritic analyses with those of terrestrial sedimentary samples for which independent exposure histories are available. As they explained: "because of the limited sensitivity of the time-integrated GCR signals provided by meteorites, it is wise to consider … also the differential GCR flux signals provided by terrestrial sediment samples" Wieler et al. (2013). This is potentially a much more powerful approach, but it relies on a terrestrial sedimentary record of cosmogenic isotopes within only the last few million years (for which ice and/or ocean sediment cores preserve an independent control of the age and depositional environment). Owing to the complexity of Earth's geological and erosional history, and the fact that Earth's atmosphere and magnetic field act to attenuate the primary GCR flux, it seems unlikely that reliance on terrestrial records alone, or a combination of terrestrial and meteoritic records, will be sufficient to measure variations in the GCR flux over the hundreds of millions of years required to reconstruct the past galactic environment of the Solar System.

It is in this context that the lunar geological record has the potential to help, mainly by providing GCR records from independently dated materials with known exposure histories spanning most of Solar System history. It is true that the lunar samples currently available for study (principally obtained by the Apollo and Luna missions of forty years ago, supplemented more recently by lunar meteorites) suffer from essentially the same problems as the meteoritic samples owing to their poorly constrained exposure histories. Indeed, the fact that the present surficial regolith (Fig. 2), from which the samples were collected, has been subject to comminution and overturning ('gardening') by meteorite impacts for the last three to four Gyr, with the result that any given sample has a poorly constrained history of burial and exhumation (and thus an unknown time-variable shielding from GCRs) greatly complicates retrieval and interpretation of the GCR record (e.g. Lorenzetti et al., 2005; Levine et al., 2007).

However, the Moon's near surface environment hosts geological formations that are far better suited to the preservation of GCRs, and indeed other evidence for the Solar System's galactic environment, than the surficial regolith sampled by previous space missions. From the point of view of obtaining ancient records of the Solar System's cosmic environment, what is required is access to materials that have been exposed to the space environment at known times and for known durations throughout Solar System history. Unlike the geologically inert asteroids, from which meteorites are mostly derived, the Moon has hosted a range of geological processes which can provide just such a temporally calibrated record by covering over, and thereby preserving, previously exposed surfaces. A key concept in this respect is that of a *palaeoregolith* – a term coined, by analogy with palaeosols in terrestrial geology, for a once surficial regolith layer that has been buried by later processes (e.g. lava flows, pyroclastic deposits, or impact crater ejecta; see, e.g., Spudis, 1996; McKay, 2009; Crawford et al., 2007, 2010). Such palaeoregoliths will preserve a record of everything that impinged upon them, including GCRs and other other aspects of the space environment, dating from their time on the surface. If the strata underlying and overlying the palaeoregolith layer can be independently dated then the record retained in the paleoregolith can be assigned a precise age and duration. In essence, this extends the argument made by Weiler et al. (2013) for the use of the terrestrial sedimentary record in interpreting variations in GCR flux to a lunar 'sedimentary' record of greatly improved fidelity and vastly longer duration.

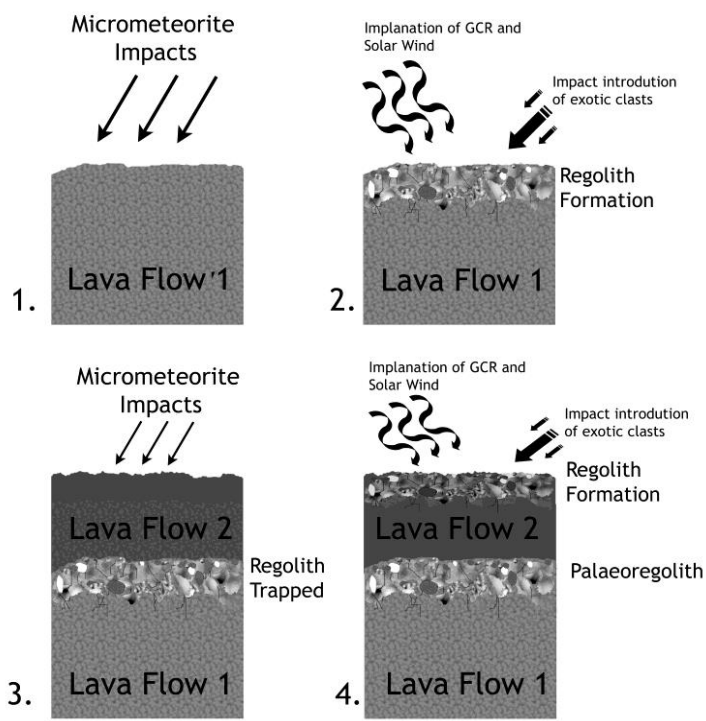

**Figure 3.** Schematic representation of the formation of a palaeoregolith layer (Crawford et al., 2007; Crawford and Joy, 2014): (1) a new lava flow is emplaced, and meteorite impacts begin to develop a surficial regolith; (2) solar wind particles, galactic cosmic ray particles and "exotic" material (possibly including supernova ejecta) are implanted; (3) the regolith layer, with its embedded historical record, is buried by a more recent lava flow, forming a palaeoregolith; (4) the process begins again on the upper surface. (Image credit: Royal Astronomical Society/K.H. Joy).

Figure 3 illustrates the basic concept in terms of palaeoregolith layers trapped between lava flows. For reasons discussed by Crawford et al. (2007, 2010), such situations will have been common on the Moon throughout the entire duration of active lunar volcanism, believed on the basis of sample studies and crater counting to have occupied at least the period 4.3 to ~1.2 Gyr ago (e.g. Hiesinger et al, 2011; Joy and Arai, 2013). Moreover, recent high-resolution images of the lunar surface have provided evidence for very young (<100 Myr), albeit small scale, volcanic activity on the Moon (Braden et al., 2014), raising the possibility that buried palaeoregolith deposits, and the record they contain, will span essentially the entire history of the Solar System. As radiometric dating of lunar basalts has now achieved a precision of a few Myr even for samples several Gyr old (e.g. Rasmussen et al., 2008), such a record should have the temporal precision to resolve changes in the galactic environment on the scale of spiral arm crossings, and attendant nearby supernova events, for multiple revolutions of the Galaxy.

Actually, as noted by Rumpf et al. (2013), if we are solely interested in the GCR record, then a palaeoregolith *per se* isn't strictly required: GCRs will create cosmogenic nuclides within the uppermost metre of so of an exposed surface whether this covered by a regolith layer or not. The required measurements could therefore be made by sampling the lava flow stratigraphy (see Fig. 4 for an example) in the absence of intervening palaeoregolith layers. That said, the insulating properties of even thin palaeoregoliths between lava flows would reduce the potential for thermal mobilization of cosmogenic nuclides due to heating from the overlying lava during its emplacement (Rumpf et al., 2013). Moreover, a well-developed surficial regolith (later palaeoregolith) layer would be desirable for the retention and preservation of other lunar astrophysical records likely to be of interest (e.g. solar wind particles, interstellar pickup ions, and interstellar dust particles).

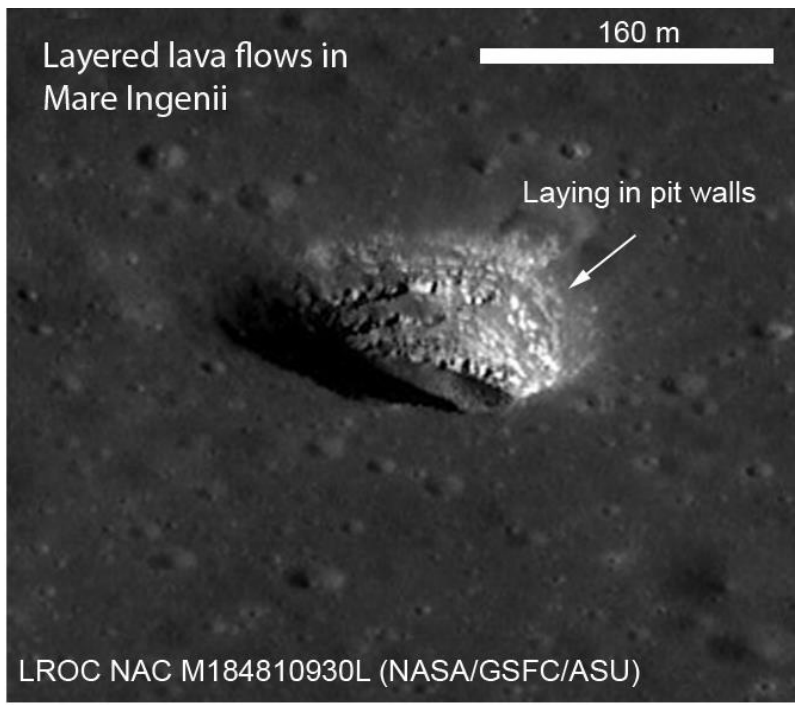

**Figure 4.** Oblique Lunar Reconnaissance Orbiter Camera view of a lunar collapse pit in Mare Ingenii, clearly revealing stratification of the basaltic lava flows in this region of the Moon. The observed layering indicates that the lunar maria are built up from multiple lava flows (Robinson et al., 2012), and that sub-surface flows that were once exposed to the surface environment, and between which palaeoregoliths may be trapped, are likely to be common on the Moon. (Crawford and Joy, 2014; NASA/GSFC/ASU).

For completeness, we note that while lava flows, and palaeoregoliths trapped between them, would be ideal for this purpose, mainly owing to the relative ease with which basalts can be radiometrically dated, other geological processes have operated on the Moon which would also tend to bury pre-existing surfaces and therefore preserve a similar record. One possibility, probably more appropriate for the older end of the record, would be the covering of previously exposed surfaces by volcanic pyroclastic deposits rather than by lava flows (McKay, 2009). Another, which in principle would cover the entire history of the Moon, but which would probably be most valuable for the younger end of the record, would be the covering of previously exposed surfaces by the ejecta blankets of craters whose ages could be determined independently of cosmic ray exposure (e.g. through the radiometric dating of associated impact melt fragments).

3.2 Supernova ejecta on the Moon

In addition to an enhanced GCR flux, it has long been argued (e.g. Benítez et al., 2002; Knie et al., 1999, 2004; Athanassiadou and Fields, 2013; Fry et al., 2015) that supernovae occurring within a few tens of pc, and perhaps $\geq$ 100 pc, of the Sun could deposit ejecta enriched in radioactive isotopes in the Solar System. Of necessity, most attempts to identify SN ejecta in the Solar System have been made on Earth, with most attention focussed on an apparent enhancement of the relatively short-lived isotope $^{60}$Fe in ocean sediments (Knie et al., 1999, 2004; Fitoussi et al., 2008; Wallner et al., 2016). In their two papers, Knie et al. (1999, 2004) reported spikes in the $^{60}$Fe concentration in two separate deep sea ferromanganese crusts which they interpreted as ejecta from a nearby SN. The study by Knie et al. (2004) had much better time resolution and, allowing for a more recent revision of the half-life of $^{60}$Fe, implies arrival of $^{60}$Fe-bearing SN ejecta, probably in the form of sub-micron-sized dust grains, in the inner Solar System ~2.2 Myr ago (Fry et al., 2015). This result has recently been confirmed by a much more extensive, global, study by Wallner et al. (2016), who identified two peaks in $^{60}$Fe deposition in ocean sediments in the age ranges 1.5-3.2 and 6.5-8.7 Myr ago. Both the inferred $^{60}$Fe flux and the timescale appear consistent with an origin in the Sco-Cen OB Association, as suggested by Benítez et al. (2002) and Breitschwerdt et al. (2016). Although Basu et al (2007) had argued that the $^{60}$Fe enhancements reported by Knie et al. (1999, 2004) may have been cosmogenic $^{60}$Fe produced in iron micrometeorites before incorporation into the ocean sediments, Fitoussi et al. (2008) argued persuasively against this re-interpretation and the recent studies by Wallner et al. (2016) and Breitschwerdt et al. (2016) very strongly support a supernova origin.

Cook et al. (2009) have argued that the lunar surface would have some advantages as a collector of SN debris because the 'sedimentation' rate, which on the Moon really equates to the regolith formation rate (~ 1mm Myr$^{-1}$ at the present epoch; McKay et al., 1991) is so low that the cosmic signal would be less diluted than it is in terrestrial sediments, and that the low concentration of Ni in lunar soils would minimise the cosmogenic (spallation) production of $^{60}$Fe. On the other hand, both the continual 'gardening' of the surface regolith by meteorite impacts, and possible meteoritic contamination itself, need to be allowed for. Cook et al. (2009) and Fimiani et al. (2012, 2014, 2016) were able to allow for both the latter effects and have successfully identified $^{60}$Fe enhancements in a number of Apollo samples which are consistent with the accretion of SN ejecta ~2 Myr ago. Although the $^{60}$Fe concentration reported for these lunar samples is about an order of magnitude less than would be expected from the $^{60}$Fe fluence estimates of Fitoussi et al. (2008) based on their interpretation of the terrestrial measurements, a number of plausible explanations for the discrepancy have been identified (Cook et al., 2009; Fry et al., 2015; Fimiani et al. 2016).

Clearly these results and their interpretation still need to be confirmed. However, even if the $^{60}$Fe found in Apollo soil samples is confirmed to be ejecta from a nearby SN ~2 Myr ago, the current surficial lunar regolith, which is all that was sampled by the Apollo and Luna missions, will not lend itself to the detection of debris from older SN events which would be more relevant to reconstructing the past galactic environment of the Solar System. In part this is because of the relatively short half-lives of isotopes likely to trace SN ejecta: Fry et al. (2015) have identified ten such isotopes, but most are similar to $^{60}$Fe in having half-lives of the order of a few Myr or less. However, in principle, a SN signature might be inferred from a careful study of the decay products of these isotopes, so live isotopes might not be required. Moreover, Fry et al. (2015) identified two radioactive isotopes, $^{146}$Sm ($T_{1/2}$ = 100 Myr) and $^{244}$Pu ($T_{1/2}$=80 My), which in principle might be able to record SN ejecta from multiple spiral arm passages.

More serious is the continual 'gardening' of the lunar surface by meteorite impacts, with the result that, even if multiple SN events have deposited long half-life isotopes over the last several hundred Myrs, the surface layers of the currently exposed regolith will only record a crude average over this time. Such considerations led Fry et al. (2015) to downplay the value of the lunar surface as a detector of SN ejecta. However, this is exactly the same problem we encountered in Section 3.1 with regard to the preservation of time-resolved GCR records on the Moon, and it has essentially the same solution. Throughout most of lunar history, any global deposition of SN ejecta on the lunar surface will necessarily have deposited material onto surface regoliths destined to be covered over by later processes, such as lava flows or crater ejecta blankets. Thus, just as for GCR records, palaeoregolith layers (Fig. 3), the ages of which can well constrained by dating the over- and under-lying strata, will be the ideal materials to search for time-resolved SN ejecta on the Moon.

Fry et al. (2015) have identified another potential problem with the identification of SN ejecta on the lunar surface: sub-micron-sized grains originating from nearby supernovae would be expected to strike the lunar surface at speeds of $\geq$100 kms$^{-1}$, resulting in their almost complete vaporisation. If most of this vapour escapes to space, rather than re-condensing in the target regolith, then only a small fraction of the SN ejecta may be retained. Indeed, Fry et al. invoked this effect to account for the lower than expected concentrations of $^{60}$Fe found by Cook et al. (2009) compared to what would have been expected based on the terrestrial detections by Knie et al. (1999, 2004) and Fitoussi et al. (2008). However, on a micron scale the surface of lunar regolith is very porous (e.g. McKay et al., 1991), which may tend to entrap escaping vapour and facilitate its condensation locally. Moreover, for a given SN event there will always be many places on the Moon where the incoming material will impact the surface at a shallow angle, which may tend to minimise shock heating of the impacting particles (e.g. Crawford et al., 2008). Clearly the effect of impact vaporisation of high-velocity SN ejecta identified by Fry et al. (2015) is potentially important, but it will require more detailed modelling of the ejecta-surface interaction before concluding that the lunar surface will be an inefficient collector of SN ejecta.

Finally, we note that if SN ejecta can be identified in the Solar System, Fry et al (2015) have shown that the fluxes and ratios of different isotopes could be used to identify the *type* of SN event responsible (see their Table 1). Clearly, if time-resolved records of multiple SN events could be identified, such as might be provided by lunar palaeoregolith deposits having a wide range of ages, this would yield very interesting astrophysical information regarding the relative frequency of different classes of supernovae encountered by the Solar System in its journey around the Galaxy.

## 4. Accessing the lunar record

Although I have argued above that the lunar geological record is likely to preserve a rich archive of the Solar System's past galactic environments, including, but not limited to, a record of nearby SN explosions, it is clear that this archive will not be easy to access. In particular, although they are the only samples we currently have access to, the Apollo and Luna samples of the present surficial regolith are not well-suited to this purpose. Much better records are likely to be preserved in buried palaeoregoliths. However, by definition, palaeoregolith deposits will be located below the present surface and a significant exploratory effort will be required to locate and sample them.

Geological mapping of lunar lava flows (e.g. Hiesinger et al., 2011), pyroclastic deposits (e.g. Gaddis et al., 2003), and the ejecta blankets of small Copernican-aged craters (which are ubiquitous on the Moon) will all help identify areas where such deposits are likely to be found. High-resolution images, such as those obtained with the Lunar Reconnaissance Orbiter Camera (e.g. Robinson et al., 2012) will be especially valuable in this respect. Studies of areas where small impact craters have excavated through overlying lava flows to reveal sub-surface boundaries (e.g. Weider et al., 2010), and/or orbital ground penetrating radar measurements (e.g. Ono et al., 2009), will help confirm the likely occurrence of palaeoregolith layers and provide some information on their depth and thickness. Ultimately, we will need to identify specific localities where buried palaeoregoliths may be accessible to future robotic or human exploration. Ideally it would be desirable to identify localities where palaeoregolith layers outcrop at the surface (e.g. in the walls of craters or rilles) or where modest drilling (e.g. to depths of a few tens of meters) would permit access to sub-surface palaeoregolith deposits. Rover-borne ground penetrating radar instruments, such as that recently successfully deployed on the Chang'e-3 rover Yutu (Long et al, 2015), would be useful in identifying the latter.

Although initial exploratory sampling might be achieved robotically, there is an argument (e.g. Spudis, 1996; Crawford and Joy, 2014) that gaining full access to lunar palaeoregolith deposits, and other scientifically valuable sites and samples on the Moon, would be greatly facilitated by a renewed human presence on the lunar surface. The requirements of an exploration architecture able to locate and access these materials has been outlined by Crawford et al. (2007, 2010) and, briefly, would consist of the following elements:

- The ability to conduct 'sortie-class' expeditions to a range of localities on the lunar surface;

- Provision for surface mobility, ideally with a range of several hundred km;

- Provision of the means to detect sub-surface palaeoregolith deposits such as a rover-borne ground penetrating radar system (e.g. Heggy et al., 2009; Long et al., 2015);

- Provision of a drilling capability, perhaps to ~ 100 m depths (for a review of suitable planetary drilling technology see Zacny et al., 2006); and

- Adequate provision for sample collection and Earth-return capacity. This clearly depends on the number and types of sites (e.g. palaeoregolith layers) sampled by each exploration mission, but based on the analysis of Shearer et al. (2007) this is roughly estimated at several 100 kg per sortie (which, for context, may be compared with the 110 kg returned from the Apollo 17 landing site in 1972) .

Clearly all this would be a very considerable undertaking, and one that will not be taken solely, or even mainly, with the aim of identifying records of past supernovae in the solar vicinity. Rather it has to be seen in the context of future lunar exploration efforts aimed at addressing a wide range of scientific, and conceivably also commercial, interests in the Moon. As reviewed by, among others, Spudis (1996), Crawford and Joy (2014) and Crotts (2014), such an expanded lunar exploration programme would have the potential to yield multiple benefits over a wide range of scientific disciplines.

Recent developments in international space policy augur well for the initiation of such an expanded lunar exploration programme. In particular, in 2013 twelve of the world's space agencies produced the Global Exploration Roadmap (ISECG, 2013), which outlines possible international contributions to the human and robotic exploration of the inner Solar System over the next twenty-five years. Implementation of the Global Exploration Roadmap would lay the foundations for a exploration programme on the scale required to exploit the lunar geological record to address the scientific objectives discussed above.

## Conclusions

In this chapter I have argued that the lunar surface, including the near sub-surface, will contain a rich record of the Solar System's changing galactic environment, including (but not limited to) records of nearby SN explosions. In particular, the lunar geological record is expected to preserve evidence for increases in the flux of GCRs resulting from nearby supernovae, as well as the direct collection of supernova ejecta. However, owing to the continual 'gardening' of the lunar surface by meteorite impacts, the present surficial regolith, which is the only part of the lunar surface that has been sampled by space missions to-date, is far from ideal for this purpose. Rather, in order to fully leverage the lunar geological record to address questions related to the changing galactic environment of the Solar System, samples will need to be collected from currently sub-surface strata (e.g. palaeoregolith layers) that were directly exposed to the space environment at known times in the past and for known durations. Such geological records certainly exist on the Moon, but accessing them will require a greatly expanded programme of lunar exploration. Such an expanded exploration programme is currently under discussion by the world's space agencies (ISECG, 2013), and its implementation would be expected to yield many scientific benefits across multiple disciplines (e.g. Spudis, 1996; Crawford and Joy, 2014; Crotts, 2014). Among the specifically astrophysical sub-set of benefits resulting from future lunar exploration will be an improved understanding of the interaction between the Solar System and its changing galactic environment, including the frequency with which the Solar System, and thus the Earth, has been exposed to nearby supernova events.

## Acknowledgements

I thank my colleagues Louise Alexander, Katherine Joy, Joshua Snape, and Pieter Vermeesch for many helpful conversations on topics related to this chapter, and The Leverhulme Trust for supporting related research activities.